# Biff (Bloom Filter) Codes:
# Fast Error Correction for Large Data Sets

Michael Mitzenmacher* and George Varghese†
*Harvard University, School of Engineering and Applied Sciences
†UCSD CSE and Yahoo Labs

*Abstract*—Large data sets are increasingly common in cloud and virtualized environments. For example, transfers of multiple gigabytes are commonplace, as are replicated blocks of such sizes. There is a need for fast error-correction or data reconciliation in such settings even when the expected number of errors is small.

Motivated by such cloud reconciliation problems, we consider error-correction schemes designed for large data, after explaining why previous approaches appear unsuitable. We introduce Biff codes, which are based on Bloom filters and are designed for large data. For Biff codes with a message of length $L$ and $E$ errors, the encoding time is $O(L)$, decoding time is $O(L+E)$ and the space overhead is $O(E)$. Biff codes are low-density parity-check codes; they are similar to Tornado codes, but are designed for errors instead of erasures. Further, Biff codes are designed to be very simple, removing any explicit graph structures and based entirely on hash tables. We derive Biff codes by a simple reduction from a set reconciliation algorithm for a recently developed data structure, invertible Bloom lookup tables. While the underlying theory is extremely simple, what makes this code especially attractive is the ease with which it can be implemented and the speed of decoding. We present results from a prototype implementation that decodes messages of 1 million words with thousands of errors in well under a second.

## I. INTRODUCTION

Motivated by the frequent need to transfer and reconcile large data sets in virtualized and cloud environments, we provide a very simple and fast error-correcting code designed for very large data streams. For example, consider the specific problem of reconciling two memories of 2 Gbytes whose contents may differ by a small number of 32-bit words. Alternatively, one can picture transferring a memory of this size, and needing to check for errors after it is written to the new storage. We assume errors are mutation errors; data order remains intact.

Other possible applications include deduplication, as exemplified by the Difference Engine [6]. While storage may seem cheap, great cost savings can be effected by replacing redundant copies of data with a single copy and pointers in other locations. For example, in virtualized environments, it is not surprising that two virtual machines might have virtual memories with a great deal of redundancy. For example, both VMs may include similar copies of the operating system. More generally, we are concerned with any setting with large data transfers over networks.

In this setting, our primary notion of efficiency differs somewhat from standard coding. While we still want the redundancy added for the code to be as small as possible, speed appears to be a more important criterion for large data sets. In particular, for a message of length $L$ and $E$ errors, while we may want close to the minimum overhead of $E$ words, $O(E)$ words with a reasonably small constant should suffice. More importantly, we require very fast encoding and decoding times; encoding should be $O(L)$ and decoding should be $O(L+E)$, with very small constant factors implied in the asymptotic notation. Typically, $E$ will be very small compared to $L$; we expect very small error rates, or even subconstant error rates (such as a bounded number of errors).

In this paper, we describe new codes that are designed for large data. We also show why other approaches (such as Reed-Solomon Codes or Tornado codes with with block-based checksum) are unsuitable. Our codes are extremely attractive from the point of view of engineering effectiveness: our software prototype implementation is very fast, decoding messages of 1 million words with thousands of errors in under a second.

We call our codes Biff codes, where Biff denotes how we pronounce BF, for Bloom filter. Biff codes are motivated by recent Bloom filter variations, the invertible Bloom filter [3] and invertible Bloom lookup table [5], and their uses for set reconciliation [4], as explained below. Alternatively, Biff codes are similar to Tornado codes [1], [9], and can be viewed as a practical, randomized low-density parity-check code with an especially simple structure designed specifically for word-level mutation errors. Also, while Tornado codes were designed using multiple levels of random graphs with carefully chosen degree distributions, Biff codes reduce this structure to its barest elements; our basic structure is single-layer, and regular, in that each message symbol takes part in the same number of encoded symbols. As a result, programming efficient encoding and decoding routines can easily be done in a matter of hours. We expect this simplicity will be prized as a virtue in practical settings; indeed, we believe Biff codes reflect the essential ideas behind other related low-density parity-check (LDPC) codes, in their simplest form.

We also provide a simple (and apparently new) general reduction from error correcting codes to set reconciliation. While reductions from erasure and error correcting codes to set reconciliation are well known [7], [10], our reduction may be useful independent of Biff Codes.

Finally, a related approach for reconciliation, focused on the setting of data streams, appears in [14].

## II. FROM SET RECONCILIATION TO ERROR CORRECTING OMISSION ERRORS

We now describe how to construct Biff codes from invertible Bloom lookup tables (IBLTs). The source of the stream of ideas we exploit is a seminal paper called Invertible Bloom

Filters by Eppstein and Goodrich that invented a streaming data structure for the so-called straggler problem [3]. The basic idea was generalized for set reconciliation by Eppstein, Goodrich, Uyeda, and Varghese in [4] and generalized and improved further by Goodrich and Mitzenmacher to IBLTs [5]. We choose to use the framework of IBLTs in the exposition that follows though we could have used the set difference algorithm in [4] instead.

We start by reviewing the main aspects of IBLTs that we require from [5]. We note that we do not require the full IBLT structure for our application, so we discuss only the elements that we need, and refer readers to [5] for further details on IBLT performance.

*A. IBLTs via Hashing*

Our IBLT construction uses a table $T$ of $m$ cells, and a set of $k$ random hash functions, $h_1, h_2, \ldots, h_k$, to store a collection of key-value pairs. In our setting, keys will be distinct, and each key will have a value determined by the key. On an insertion, each key-value pair is placed into cells $T[h_1(x)]$, $T[h_2(x)]$, $\ldots T[h_k(x)]$. We assume the hash functions are fully random (hash values independent and uniformly distributed); in practice this assumption appears suitable (see, e.g., [11], [13] for related work on this point). For technical reasons, we assume that distinct hash functions yield distinct locations. This can be accomplished in various ways, such as by splitting the $m$ cells into $k$ subtables each of size $m/k$, and having each hash function choose one cell (uniformly) from each subtable. Such splitting does not affect the asymptotic behavior in our analysis.

In a standard IBLT, each cell contains three fields: a keySum field, which is the exclusive-or (XOR) of all the keys that have been inserted that map to this cell; a valueSum field, which is the XOR of all the values of the keys that have been inserted that map to this cell; and a count field, which counts the number of keys that have been inserted into the cell.

As all operations are XORs, deletions are handled in an equivalent manner: on deletion of a previously inserted key-value pair, the IBLT XORs the key and value with the fields in the appropriate cells, and the count is reversed. This reverses a corresponding insertion. We will discuss later how to deal with deletions without corresponding insertions, a case that can usefully occur in our setting.

*B. Listing Set Entries*

We now consider how to list the entries of the IBLT. The approach is straightforward. We do a first pass through the cells to find cells with a count of 1, and construct a list of those cells. We recover the key and corresponding value from this cell, and then delete the corresponding pair from the table. In the course of performing deletions, we check the count of the relevant cells. If a cell's count becomes 1, we add it to the list; if it drops from 1 to 0, we can remove it from the list. This approach can easily be implemented $O(m)$ time.

If at the end of this process all the cells have a count of 0, then we have succeeded in recovering all the entries in the IBLT. Otherwise, the method only outputs a partial list of the key-value pairs in $\mathcal{B}$.

| $k$ | 3 | 4 | 5 | 6 | 7 |
|---|---|---|---|---|---|
| $c_k$ | 1.222 | 1.295 | 1.425 | 1.570 | 1.721 |

TABLE I
THRESHOLDS FOR THE 2-CORE ROUNDED TO FOUR DECIMAL PLACES.

This "peeling process" is well known in the context of random graphs and hypergraphs as the process used to find the 2-core of a random hypergraph (e.g., see [2], [12]). This peeling process is similarly used for various codes, including Tornado codes and their derivatives (e.g., see [9]). Previous results therefore give tight thresholds: when the number of hash values $k$ for each pair is at least 2 and there are $n$ key-value pairs in the IBLT, there are constants $c_k > 1$ such that if $m > (c_k + \epsilon)n$ for any constant $\epsilon > 0$, the listing process succeeds with probability $1-o(1)$; similarly, if $m < (c_k-\epsilon)n$ for any constant $\epsilon > 0$, the listing process fails with probability $1-o(1)$. As shown in [2], [12], these values are given by

$$c_k^{-1} = \sup\left\{\alpha : 0 < \alpha < 1; \forall x \in (0,1), 1 - e^{-k\alpha x^{k-1}} < x\right\}.$$

Numerical values for $k \geq 3$ are given in Table I.

The choice of $k$ affects the probability of the listing process failing. By choosing $k$ sufficiently large and $m$ above the 2-core threshold, standard results give that the bottleneck is the possibility of having two key-value pairs with the same collection of hash values, giving a failure probability of $O(m^{-k+2})$.

We note that, with some additional effort, there are various ways to save space with the IBLT structure that are known in the literature. including using compressed arrays, quotienting, and irregular constructions (where different keys can utilize a different number of hash values, as in irregular LDPC codes). In practice the constant factors are small, and such approaches may interfere with the simplicity we aim for with the IBLT approach; we therefore do not consider them further here.

*C. Set Reconciliation with IBLTs*

We consider two users, Alice and Bob, referred to as $A$ and $B$. Suppose Alice and Bob hold distinct but similar sets of keys, and they would like to reconcile the differences. This is the well known set reconciliation problem. To achieve such a reconciliation with low overhead [4], Alice constructs an IBLT. The value associated with each key is a fingerprint (or checksum) obtained from the key. In what follows, we assume the value is taken by hashing the key, yielding an uniform value over all $b$-bit values for an appropriate $b$, and that the hash function is shared between Alice and Bob. Alice sends Bob her IBLT, and he correspondingly deletes from the IBLT the key-value pairs from his set.

In this setting, when Bob deletes a key-value pair not held by Alice, it is possible for a cell count to become negative. The remaining key-value pairs left in the IBLT correspond exactly to items that exactly one of Alice or Bob has. Bob can use the IBLT structure to recover these pairs efficiently. For lack of space, we present the argument informally; the IBLT properties we use were formally derived in [5].

We first note that, in this setting, because deletions may reduce the count of cells, it is possible that a cell can have a count of 1 but not contain exactly one key-value pair. For example, if two pairs are inserted by Alice into a cell, and Bob deletes a pair that does not match Alice's in that cell, the count will be 1. Hence, in this setting, the proper way to check if a cell contains a valid pair is to test that the checksum for the keySum field matches the valueSum. In fact, because the value corresponds to a checksum, the count field is extraneous. (It can be a useful additional check, but strictly is unnecessary. Moreover, it may not be space-effective; since the counts will depend on the number of items inserted, not on the size of the difference between the sets.) Instead, the list of cells that allow us to recover a value in our listing process are determined by a match of the key and checksum value. Importantly, because Bob's deletion operation is symmetric to Alice's insertion operation, this holds true for cells containing a pair deleted by Bob as well as cells containing a pair inserted by Alice. (In this case, the corresponding count, if used, should be $-1$ for cells with a deleted pair.)

Bob can therefore use the IBLT to recover these pairs efficiently. (Strictly speaking, Bob need only recover Alice's keys, but this simplification does not make a noticeable difference in our context.) If $\Delta$ is an upper bound on the number of keys not shared between Alice and Bob, then from the argument sketched above, an IBLT with only $O(\Delta)$ cells is necessary, with the constant factor dependent on the success probability desired.

## III. Error-Correcting Codes with IBLTs

We now show how to use the above scheme to obtain a computationally efficient error-correcting code. Our error-correcting code can be viewed as a reduction using set reconciliation. Let $B$ have a message for $A$ corresponding to the sequence of values $x_1, x_2, \ldots, x_n$. Then $B$ sends $A$ the message along with set reconciliation information – in our case, the IBLT – for the set of ordered pairs $(x_1, 1), (x_2, 2), \ldots, (x_n, n)$. For now we assume the set reconciliation information $A$ obtains is without error; errors only occur in message values. When $A$ obtains the sequence $y_1, y_2, \ldots, y_n$, she constructs her own set of pairs $(y_1, 1), (y_2, 2), \ldots, (y_n, n)$, and reconciles the two sets to find erroneous positions. Notice that this approach requires random symbol errors as opposed to adversarial errors for our IBLT approach, as we require the checksums to accurately determine when key-value pairs are valid. However, there are standard approaches that overcome this problem that would make it suitable for adversarial errors with a suitably limited adversary (by applying a pseudo-random permutation on the symbols that is secret from the adversary; see, for example, [8]). Also, the positions of the errors can be anywhere in the message (as long as the positions are chosen independently of the method used to generate the set reconciliation information).

If there are no errors in the data for the IBLT structure, then this reduction can be directly applied. However, assuming the IBLT is sent over the same channel as the data, then some cells in the IBLT will have erroneous keySum or valueSum fields. If errors are randomly distributed and the error rate is sufficiently small, this is not a concern; as shown in [5], IBLT listing is quite robust against errors in the IBLT structure. Specifically, an error will cause the keySum and valueSum fields of an IBLT cell not to match, and as such it will not be used for decoding; this can be problematic if all the cells hashed to an erroneous message cell are themselves in error, as the value cannot then be recovered, but under appropriate parameter settings this will be rare in practice. As a summary, using the 1-layer scheme, where errors can occur in the IBLT, the main contribution to the failure probability is when an erroneous symbol suffers from all $k$ of its hash locations in the IBLT being in error. If $z$ is the fraction of IBLT cells in error, the expected number of such symbols is $Ez^k$, and the distribution of such failures is binomial (and approximately Poisson, when the expectation is small). Hence, when such errors occur, there is usually only one of them, and instead of using recursive error correction on the IBLT one could instead use a very small amount of error correction in the original message.

For bursty errors or other error models, we may need to randomly intersperse the IBLT structure with the original message; note, however, that the randomness used in hashing the message values protects us from bursty errors over the message.

Basic pseudocode for encoding and decoding of Biff codes is given below (using C-style notation in places); the code is very simple, and is written entirely in terms of hash table operations.

- ENCODE
    **for** $i = 1 \ldots n$ **do**
        **for** $j = 1 \ldots k$ **do**
            $T_j[h_j((x_i, i))]$.keySum $\hat{=} (x_i, i)$.
            $T_j[h_j((x_i, i))]$.valueSum $\hat{=}$ Check$((x_i, i))$.
- DECODE
    **for** $i = 1 \ldots n$ **do**
        **for** $j = 1 \ldots k$ **do**
            $T_j[h_j((y_i, i))]$.keySum $\hat{=} (y_i, i)$.
            $T_j[h_j((y_i, i))]$.valueSum $\hat{=}$ Check$((y_i, i))$.
    **while** $\exists\ a, j$ with $(T_j[a]$.keySum $\neq 0)$ and $(T_j[a]$.valueSum $==$ Check$(T_j[a]$.keySum$))$ **do**
        $(z, i) = T_j[a]$.keySum
        **if** $z \neq y_i$ **then**
            set $y_i$ to $z$ when decoding terminates
        **for** $j = 1 \ldots k$ **do**
            $T_j[h_j((z, i))]$.keySum $\hat{=} (z, i)$.
            $T_j[h_j((z, i))]$.valueSum $\hat{=}$ Check$((z, i))$.

In our pseudocode, there is some leeway in how one implements the while statement. One natural implementation would keep a list (such as a linked list) of pairs $a, j$ that satisfy the conditions. This list can be initialized by a walk through the arrays, and then updated as the while loop modifies the contents of the table. The total work will clearly be proportional to the size of the tables, which will be $O(E)$ when the table size is chosen appropriately.

We may also recursively apply a further IBLT, treating the first IBLT as data, or we can use a more expensive error-correcting code, such as a Reed-Solomon code, to protect the much smaller IBLT. This approach is similar to that used under the original scheme for Tornado codes, but appears unnecessary for many natural error models. For ease of exposition, we assume random locations for errors henceforth.

The resulting error-correcting code is not space-optimal, but the overhead in terms of the space required for the error-correction information is small when the error-rate is small. If there are $e$ errors, then there will be $2e$ key-value pairs in the IBLT; the overhead with having 3, 4, or 5 choices, as seen from Table I, will then correspond to less than $3e$ cells. Each cell contains both a keySum or valueSum, each of which will be (depending on the implementation) roughly the same size as the original key. Note here the key in our setting includes a position as well as the original message symbol, so this is additional overhead. Putting these together, we can expect that the error-correction overhead is roughly a factor of 6 over the optimal amount of overhead, which would be $e$ times the size of a message symbol.

While this is a non-trivial price, it is important to place it in context. For large keys, with a 1% error rate, even an optimal code for a message of length $M$ bytes would require at least $(1/0.99)M \approx 1.01M$ bytes to be sent, and a standard Reed-Solomon code (correcting $E$ errors with $2E$ additional values) would require at least $1.02M$ bytes. Biff codes would require about $1.06M$ bytes. The resulting advantages, again, are simplicity and speed. We expect that in many engineering contexts, the advantages of the IBLT approach will outweigh the small additional space cost.

For very large messages, parallelization can speed things up further; key-value pairs can be inserted or deleted in parallel easily, with the bottleneck being atomic writes when XORing into a cell. The listing step also offers opportunities for parallelization, with threads being based on cells, and cells becoming active when their checksum value matches the key. We don't explore parallelization further here, but we note the simple, regular framework at the heart of Biff codes.

We also note that, naturally, the approach of using IBLTs can be applied to design a simple erasure-correcting code. This corresponds to a set reconciliation problem where one set is slightly larger than the other; nothing is inserted at A's end for missing elements. Other error models may also be handled using the same technique.

## IV. Issues with Other Approaches

Other natural approaches fail to have both fast encoding and decoding, and maintain $O(E)$ overhead. While asymptotically faster algorithms exist, the computational overhead of Reed-Solomon codes is generally $\Theta(EL)$ in practice, making a straightforward implementation infeasible in this setting, once the number of errors is non-trivial. Breaking the data into blocks and encoding each would be ineffective with bursty errors. One could randomly permute the message data before breaking it into blocks, to randomize the position of errors and thereby spread them among blocks. In practice, however, taking a large memory block and then permuting it is extremely expensive as it destroys natural data locality. Once a memory or disk page is read it is almost "free" to read the remaining words in sequence; randomizing positions becomes hugely expensive. (By contrast, the method we suggested to make Biff more resilient to adversarial errors does not destroy locality because it randomizes over *symbol values*, not *positions*.) Finally, there are issues in finding a suitable field size to compute over, particularly for large messages.

The problems we describe above are not original; similar discussions, for example, appear with the early work on Tornado codes [1]. Experiments comparing Reed-Solomon codes for erasures with Tornado codes from the original paper demonstrate that Reed-Solomon codes are orders of magnitude slower at this scale.

An alternative approach is to use Tornado codes (or similar LDPC codes) directly, using checksums to ensure that suitably sized blocks are accurate. For example, we could divide the message of length $L$ into $L/B$ blocks of $B$ symbols and add an error-detection checksum of $c$ bits to each block. If we assume blocks with detected errors are dropped, then $E$ errors could result in $EB$ symbols being dropped, requiring the code to send at least an additional $kEB$ bits for a suitably small constant $k$. The total overhead would then be $Lc/B + kEB$; simple calculus yields the minimum overhead is when $B = 2\sqrt{ckLE}$, with block sizes of $O(\sqrt{L/E})$ and resulting space overhead of $O(\sqrt{LE})$.

On the other hand, for Biff codes the redundancy overhead is $O(E)$ with small constants hidden in the $O$ notation, because only the values in the cells of the hash table, and not the original data, require checksums. This is a key benefit of the Biff code approach; only the hash table cells need to be protected with checksums.

## V. Experimental Results

In order to test our approach, we have implemented Biff codes in software. Our code uses pseudorandom hash values generated from the C drand function (randomly seeded using the clock), and therefore our timing information does not include the time to hash. However, we point out that hashing is unlikely to be a major bottleneck. For example, even if for each one wants 4 hash locations for each key into 4 subtables of size 1024, and an additional 24 bit hash for the checksum for each key, all the necessary values can be obtained with a single 64-bit hash operation.

*Setup:* Our has table is split into $k$ equal subtables. As mentioned, to determine locations in each subtable, we use pseudorandom hash values. For convenience we use random 20 bit keys as our original message symbols and 20 bits to describe the location in the sequence. While these keys are small, it allows us to do all computation with 64-bit operations. For a checksum, we use a simple invertible function: the pair $(x_i, i)$ gives a checksum of $(2i + 1) * x_i + i^2$.

One standard test case uses 1 million 20-bit message symbols and an IBLT of 30000 cells, with errors introduced in 10000 message symbols and 600 IBLT cells. Note that with only 20 bit keys and 20 bits to record the length, an IBLT cell is actually 4 times the size of a message cell (so our IBLT is 2400000 bits); however, we use a 2% error rate in the IBLT as we expect message symbols will generally be much longer. For example, in practice a key might be a 1KB packet, in which case 1 million message symbols would correspond to a gigabyte.

*Timing:* Our results show Biff codes to be extremely fast. There are two decoding stages, as can be seen in the previously given pseudocode. First, the received sequence values must be placed into the hash table. Second, the hash table must be processed and the erroneous values recovered. Generally, the bulk of the work will actually be in the first stage, when the number of errors are small. We had to utilize messages of 1 million symbols in order to obtain suitable timing data; otherwise processing was too fast. On our standard test case over 1000 trials, using 4 hash functions the first stage took 0.0561 seconds on average and the second took 0.0069 seconds on average. With 5 hash functions, the numbers were 0.0651 second and 0.0078 seconds.

*Thresholds:* Our threshold calculations are very accurate. For example, in a setting where no errors are introduced in the IBLT, with 4 hash functions and 10000 errors we would expect to require approximately 26000 cells in order to recover fully. (Recall that 10000 errors means 20000 keys are placed into the IBLT.) Our experiments yielded that with and IBLT of 26000 cells, complete recovery occurred in 803 out of 1000 trials; for 26500 cells, complete recovery occurred in 10000 out of 10000 trials.

*Failure probabilities:* We have purposely chosen parameters that would lead to failures, in order to check our analysis. Under our standard test case with four hash functions, we estimate the probability of failure during any single trial as $10000 \cdot (600/30000)^4 = 1.6 \times 10^{-3}$. Over an experiment with 10000 trials, we indeed found 16 trials with failures, and in each failure, there was just one unrecovered erroneous message symbol. Reducing to 500 errors in the IBLT reduces the failure probability to $10000 \cdot (500/30000)^4 \approx 7.7 \times 10^{-4}$; an experiment with 10000 trials led to a seven failures, each with just one unrecovered erroneous message symbol. Finally, with 5 hash functions and 600 IBLT errors, we would estimate the failure probability as $10000 \cdot (600/30000)^5 = 3.2 \times 10^{-5}$; a run of 10000 trials yielded no failures.

## VI. Conclusions

Our goal was to design an error-correcting code that would be extremely fast and simple for use in networking applications such as large-scale data transfer and reconciliation in cloud computing systems. While not optimal in terms of rate, the amount of redundancy used is a small constant factor more than optimal; we expect this will be suitable for many applications, given the other advantages. Although we have focused on error correction of large data, Biff codes may also be useful for *smaller* messages, in settings where computational efficiency is paramount and where small block sizes were introduced at least partially to reduce Reed-Solomon decoding overheads.

We note that in the large data setting we can adapt the sampling technique described in [4] to estimate the number of errors $E$ in $O(\log L)$ time with an error bound that is quantified in Theorem 2 of [4]. This allows the Biff code to be sized correctly to $O(E)$ without requiring any a priori bound on $E$ to be known in advance. For example, when two large virtual memories are to be reconciled it is difficult to have a reasonable bound on the number of errors or differences. In the communications setting this is akin to estimating the channel error *rate* and adapting the code. However, such error rate estimation in the communication setting is done infrequently to reduce overhead. In our large data setting, the cost of estimation is so cheap that it can be done on each large data reconciliation.

Finally, we note that modern low-density parity-check codes are sufficiently complex that they are difficult to teach without without going through a number of preliminaries. By contrast, Biff codes are sufficiently simple that we believe they could be taught in an introductory computer science class, and even introductory level programmers could implement them. Beyond their practical applications, Biff codes might prove worthwhile as a gateway to modern coding techniques.

**Acknowledgments:** Michael Mitzenmacher was supported by NSF grants CCF-0915922 and IIS-0964473. George Varghese was supported by NSF Grant CSR-0964395.